\documentstyle[12pt]{article}

\begin{document}
\begin{titlepage}
\begin{flushright}
IMSc - 92 -17(Rev)\\
\end{flushright}
\begin{center}
{\large {\bf {Asymptotic Analysis and Spectrum of Three Anyons}}}
\end{center}
\vspace{1cm}
\begin{center}
G.Date, M.Krishna and M.V.N.Murthy
\end{center}
\begin{center}
The Institute of Mathematical Sciences, Madras 600 113, India
\end{center}
\vspace{2cm}
\begin{abstract}
The spectrum of anyons confined in harmonic oscillator potential
shows both linear and nonlinear dependence on the statistical
parameter.  While the existence of exact linear solutions have
been shown analytically, the nonlinear dependence has been arrived
at by numerical and/or perturbative methods.  We develop a method
which shows the possibility of nonlinearly interpolating
spectrum. To be specific we analyse the eigenvalue
equation in various asymptotic
regions for the three anyon problem.
\end{abstract}
\end{titlepage}
\section{Introduction}
Non-relativistic quantum mechanics in two spatial dimensions admits
possibility of fractional statistics \cite{lm,anyonh}.  Existence of fractional
statistics is also intimately tied to having multivalued wave functions
which occur naturally in the quantum mechanics on multiply
connected spaces \cite{braid}.  The classification of multivalued wave
functions is provided by one dimensional representations of the
fundamental group of the multiply connected configuration space.
Although there are several known examples of a kinematic classification
\cite{cmp} (i.e., representations of the fundamental group), very little is
known
about the dynamics on such spaces.  A system of ``Anyons''-
particles in two
dimensions obeying fractional statistics- is a  good and physically
relevent example of quantum mechanics on multiply connected spaces.  It
is physically relevent because anyons have been proposed as candidates
for explaining fractional quantum Hall effect \cite{fqh}
and are likely candidates
for high $T_c$ superconductivity \cite{htc}.  It is a good example because at
least some exact solutions to the energy eigenvalue problem for N-anyons
in some external confining potential are known - an exception in
many-body quantum mechanics with non-seperable Hamiltonians.

Concentrating on anyons in a harmonic oscillator potential, one notices
two important features: All the exactly known energy eigenvalues are
linear in the statistical parameter $\alpha$ for arbitrary number
of anyons \cite{exact}.  Whereas in the case
of three and four anyons, numerical studies exhibit some eigenvalues with
non-linear dependence on $\alpha$ as well as level crossings \cite{numeral}.
Meanfield studies for large number
of anyons also exhibit the same feature \cite{tf}. There is no analytical
approach exhibiting such nonlinearly interpolating energy eigenvalues-
raising perhaps a doubt that the nonlinearity may be an artifact of
numerical computations which are necessarily limited by computing
power.
In the present work we develop an analytical approach which does show
the possibility of nonlinearly interpolating spectrum. We believe
that the method is general enough to have validity even beyond the
domain of the problem on hand.

Our approach is based on analysing the eigenvalue equation (differential
equation) in various
asymptotic regions.  Noting that eigenvalues themselves are independent
of where the differential equation is solved, we expect the asymptotic
form of the differential equation to determine the spectrum.  Presently
we have analysed in detail the case of three anyons in an oscillator
potential but we believe the method can be suitably adapted to many
anyons and possibly to some other many body systems.

The three anyon problem in the asymptotic regions we consider reduces
to solving a Confluent Hypergeometric Equation (CHE) in the region
$R \le x < \infty $.  This equation has a solution regular at $x=0$ and
a solution which is irregular at $x=0$.  The regular solution leads
to linearly interpolating energies while the irregular one (R is strictly
greater than zero) leads to nonlinearly interpolating solutions.  We
give two first order equations which specify the spectrum exactly.

The paper is organised as follows:  In Sec.2 we discuss the preliminaries,
specify our formulation and set the notation.  This part is well
known and has been included to make the paper self contained.  In
Sec.3, we give a quick review of known exact solutions.  We point
out how nonlinearity can come about and note additional asymptotic
regions available to extract information about the spectrum.  After
specifying the asymptotics we derive the equations determining the
spectrum.  We note how the numerically known low lying eigenvalues fit
into our equations.  Full analysis of these system of equations is
under investigation.  Section 5 contains summary and a brief discussion.

\section{Preliminaries}
In the following by ``anyons'' we mean a quantum mechanical system
of N particles moving in two dimensions with wave functions which have
a stipulated multi-valuedness.  To make this explicit let us denote
a generic wave function as $\psi (\vec{r}_1,...,\vec{r}_N)$.  Let
$P_{ij\gamma}$ denote the operation of taking the ith particle
coordinate around the jth coordinate along a closed path $\gamma$.
The path $\gamma$ does not enclose any other particle coordinate and is
taken in an anticlockwise sense, say.  Then under such an operation
$\psi$ acquires a phase namely,
\begin{equation}
P_{ij\gamma}\psi (\vec{r}_1,...\vec{r}_N) = exp(i2\pi\alpha)\psi
(\vec{r}_1,...\vec{r}_N) .
\end{equation}

If a path $\gamma$ encloses other particle coordinates as well then
such a path can be broken into a set of closed paths each of which
encloses exactly one particle coordinate.  Applying the stipulation
above, one can compute the total phase, for such a path.  If the sense
of the path is reversed then $\alpha \rightarrow -\alpha$.  Clearly
the phase acquired depends only on the homotopy class of the
path (i.e., is same for two paths $\gamma$ and $\gamma '$ if
$\gamma$ and $\gamma '$ can be continuously deformed into each other).

Let us introduce the complex notation for particle coordinates:
$Z_j = x_j +i y_j, \bar Z_j = \bar x_j -i \bar y_j$
coordinates.  Clearly $Z_{ij}$ , where $Z_{ij}=Z_i - Z_j$, has the
property that if $Z_j$ is taken around $Z_i$, $Z_{ij}^\alpha$ changes
by  $exp(i2\pi\alpha)$.  This allows us to write any generic wave
function satisfying eqn(1) as,
\begin{equation}
\psi (Z_i,\bar Z_i) = \left [ \prod_{i<j}(\frac {Z_{ij}}{\bar Z_{ij}})^
{\alpha/2} \right ]\tilde\psi (Z_i,\bar Z_i),
\end{equation}
with the bracketed expression being a phase
and now  $\tilde \psi (Z_i,\bar Z_i)$ is a single valued function.

Clearly,
\begin{equation}
\nabla_k \psi (Z_i,\bar Z_i) = \prod_{i<j}(\frac {Z_{ij}}{\bar Z_{ij}})^
{\alpha/2} \left [ \nabla_k \tilde\psi (Z_i,\bar Z_i)
+ \nabla_k ln(\prod_{i<j}(\frac {Z_{ij}}{\bar Z_{ij}})^
{\alpha/2}) \tilde\psi (Z_i,\bar Z_i) \right ].
\end{equation}
which can be rewritten as,
\begin{equation}
\nabla_k \psi (Z_i,\bar Z_i) = \prod_{i<j}(\frac {Z_{ij}}{\bar Z_{ij}})^
{\alpha/2} \left [ \nabla_k \tilde\psi (Z_i,\bar Z_i)
+ i\alpha \sum_{j\ne k}\frac{\hat z \times \vec {r}_{kj}}
{{\mid \vec {r}_{kj} \mid}^2} \tilde\psi (Z_i,\bar Z_i) \right ].
\end{equation}
Since $\tilde \psi$ is single valued, the r.h.s of the above equation has
exactly the
same multivaluedness as the l.h.s.  In otherwords we have,
\begin{equation}
\nabla_k [\prod_{i<j}(\frac {Z_{ij}}{\bar Z_{ij}})^
{\alpha/2} \tilde\psi (Z_i,\bar Z_i)] =
\prod_{i<j}(\frac {Z_{ij}}{\bar Z_{ij}})^\frac{\alpha}{2} D_k \tilde \psi,
\end{equation}
where
$$ D_k \tilde \psi = [\nabla_k + \vec A_k ]\tilde \psi $$
and
$$ A_k(\vec r_k) = i\alpha \sum_{j\ne k}
\frac{\hat z \times \vec {r}_{kj}}
{{\mid \vec {r}_{kj} \mid}^2} $$

This allows us to write any higher order differential operators on
$\psi$ in terms of corresponding covariant differential operators on
the singlevalued wave function $\tilde \psi$.  In particular a
Hamiltonian operator, typically $ - \sum_i \nabla^2_i +V$ can be
written similarly.  An eigenvalue equation written in terms of
$\psi$ can then be recast as a corresponding equation in terms of
$\tilde \psi$.

Although both formulations are equivalent, dealing with operators
on multivalued wave functions is much less transparent than
dealing with operators on singlevalued wave functions.  Naive
commutation rules, symmetries that one would expect by looking
at an operator on single valued functions are not at all true
in general for the ``same" differential operators acting on multivalued
wave function.

Considering eigenvalue problem in terms of $\tilde \psi$ has other
advantages too.  Since all the subtleties of multivaluedness are
equivalently transcribed in terms of additional ``interaction''
terms (the so called statistical interactions), the eigenvalue
problem is amenable to approximations.  One is also on firmer ground
in doing usual algebraic manipulations with operators.   With these in
mind we will work with singlevalued wave functions with ``statistical
interactions''.

As a first step one would like to understand the system of ``free
anyons''.  However, the statistical interaction falls off as
$\mid r_{ij}\mid ^{-2}$ as $\mid r_{ij} \mid \rightarrow \infty$.
So one is not sure whether the Hamiltonian with only statistical
interactionss has only discrete eigenvalues.  One can put the
system in a box to ensure discrete eigenvalues but then one needs
suitable boundary condition. An oscillator potential ensures
discrete spectrum without introducing a finite size.  One could
take some other confining potential but in the limit $\alpha \rightarrow
0$ one should know the spectrum. One then has hope of doing at least
the perturbative analysis.  Since the statistical interaction
depends only on relative separations, the Centre of Mass(CM) dynamics
should play a trivial role and oscillator potential also allows
a separation of CM and relative coordinate dynamics.  The oscillator
potential problem can also be mapped on to a problem of anyons in
a real, constant external magnetic field along the Z-axis \cite{hotta}.
Bearing these facts in mind, we choose the oscillator potential
without further justification.  In order to derive the thermodynamic
properties of a system of anyons there exist well defined methods
of eliminating the dependence on the oscillator frequency \cite{mccabe}.

The Hamiltonian we consider is- after carrying out the usual scaling
of variables $(\hbar = c = 1)$ - interms of dimensionless
quantities,

\begin{equation}
H =\hbar \omega[\frac {1}{2}\sum_{i=1}^{N}{p}_{i}^{2} +
\frac {1}{2}\sum_{i=1}^{N} r_{i}^{2}
-\alpha\sum_{j>i=1}^{N}\frac{\ell_{ij}}{r_{ij}^2}
+\frac{\alpha^{2}}{2}\sum_{i\neq j,k}^N\frac{\vec{r}_{ij}.\vec{r}_{ik}}
{r_{ij}^{2}  r_{ik}^{2}}],
\end{equation}
where
\[\ell_{ij}= (\vec{r}_{i} - \vec{r}_{j})\times(\vec{p}_{i}-\vec{p}_{j}).\]
and all distances have been expressed in units of $1/\sqrt {m\omega}$.
Notice that the statistical interaction is independent of the centre of mass.
This is the operator we analyse subsequently.

\section{Asymptotic Analysis}
First, we briefly discuss the class of exact solutions already
known and then move over to the asymptotic analysis.
For discussing the known class of exact solutions it is convenient
to use the complex coordinates $Z_i, \bar Z_i$ in terms of which
the Hamiltonian takes the form,
\begin{equation}
H = -2 \partial_i \bar \partial_i + \frac{1}{2} Z_i \bar Z_i
- \alpha \sum_{i<j}(\frac {\partial_{ij}}{\bar Z_{ij}} -
\frac {\bar \partial_{ij}}{ Z_{ij}}) +\frac {\alpha^2}{2}
\sum_{i \ne j,k} \frac {1}{\bar Z_{ij}Z_{ik}},
\end{equation}
where $\partial_i = \partial /\partial Z_i; \partial_{ij}=\partial_i
-\partial_j$, etc.,
and the eigenvalue equation is,
\begin{equation}
H\psi(Z_i,\bar Z_i) = E\psi(Z_i,\bar Z_i).
\end{equation}
The conserved angular momentum J, is given by,
\begin{equation}
J=Z_i\partial_i -\bar Z_i \bar \partial_i.
\end{equation}
For the sake of completeness we recapitulate the standard asymptotic
analysis, i.e,$Z_i=\lambda \hat Z_i, \lambda \rightarrow \infty$ :
\begin{flushleft}
(i) As $Z_i \rightarrow \infty$,   $Z_{ij} \rightarrow \infty$, the
oscillator potential term dominates over the $\alpha$ dependent terms
reducing $H$ to the Hamiltonian of N two dimensional oscillators.  All
the eigenfunctions of the oscillator have the Gaussian suppression
factor.  Thus we put,
\begin{equation}
\psi(Z_i,\bar Z_i) = exp(-\frac{1}{2}\sum_i Z_i\bar Z_i)\psi_1(Z_i,
\bar Z_i).
\end{equation}
Substitution of this form transforms the original  eigenvalue equation
to an equation for $\psi_1$, viz.,
\begin{equation}
[ -2 \partial_i \bar \partial_i +  Z_i \partial_i + \bar Z_i
 \bar \partial_i
- \alpha \sum_{i<j}(\frac {\partial_{ij}}{\bar Z_{ij}} -
\frac {\bar \partial_{ij}}{ Z_{ij}}) +\frac {\alpha^2}{2}
\sum_{i \ne j,k} \frac {1}{\bar Z_{ij}Z_{ik}}]\psi_1 =
(E-N)\psi_1 .
\end{equation}
The N on the r.h.s is the zero point energy for N two dimensional
oscillators while on the l.h.s the oscillator potential is traded
for the ``scaling operator'',$ Z_i \partial_i +
\bar Z_i \bar \partial_i $.   \\
(ii) The other asymptotic region is $\mid Z_{ij}\mid \rightarrow 0$
for any pair of particles i and j, say.  Since the $\alpha$ dependent
terms are singular we expect the wave function to vanish sufficiently
fast in this limit so that the eigenvalue equation is well defined.

Assuming a Taylor series expansion in $Z_{ij}$ as
$\mid Z_{ij}\mid \rightarrow 0$ , we can write,
\begin{equation}
\psi_1(Z_i,\bar Z_i) \;\; \longrightarrow
 \sum _{a \: \ge \:0} (\mid Z_{12}\mid)^{\lambda+a} \psi_{1a}(Z_2,...,\bar
Z_2,...)
\end{equation}
for $\lambda \ge 0$ and we have taken the pair (12) for definiteness.
Substituting this in the eigenvalue equation and keeping terms to the leading
order in $\mid Z_{12}\mid$ gives,
$$\lambda^2-\alpha^2=0$$
or $\lambda =\pm \alpha$.  However $\lambda > 0$, thus $\lambda =
\alpha$.  Since this must happen for each pair of particles seperately
and independently we can take out a factor $\mid X\mid^\alpha$, where
\begin{equation}
X = \prod_{i<j} Z_{ij}.
\end{equation}
\end{flushleft}
For future reference let us also note that if
\begin{equation}
\psi_1 = X^{\beta /2} \bar X^{\gamma /2}\psi_2,
\end{equation}
then the eigenvalue equation transforms to
\begin{eqnarray*}
 & & [\; -2 \partial_i \bar \partial_i +  Z_i \partial_i +
\bar Z_i \bar \partial_i
 - (\alpha + \gamma) \sum_{i<j}\frac {\partial_{ij}}{\bar Z_{ij}} +
(\alpha -\beta)\sum_{i<j}\frac {\bar \partial_{ij}}{ Z_{ij}} +   \\
 & & \frac {(\alpha-\beta)(\alpha+\gamma)}{2}
\sum_{i \ne j,k} \frac {1}{\bar Z_{ij}Z_{ik}} \; ] \psi_2
 = \left[ E-N-\frac{(\beta+\gamma)}{4}N(N-1) \right] \psi_2
\end{eqnarray*}
Taking $\beta = \gamma =\alpha$, we have,
\begin{equation}
\psi_1 = \mid X \mid^{\alpha }\psi_2,
\end{equation}
and
\begin{equation}
 [ -2 \partial_i \bar \partial_i +  Z_i \partial_i +
\bar Z_i \bar \partial_i
 - 2\alpha\sum_{i<j}\frac {\partial_{ij}}{\bar Z_{ij}}\:]\:\psi_2
 = [E-N-\frac {\alpha}{2}N(N-1)\:]\:\psi_2
\end{equation}
while if $\beta = \gamma =-\alpha$, we have,
\begin{equation}
\psi_1 = \mid X \mid^{-\alpha }\psi_2',
\end{equation}
and
\begin{equation}
 [ -2 \partial_i \bar \partial_i +  Z_i \partial_i +
\bar Z_i \bar \partial_i
 + 2\alpha\sum_{i<j}\frac {\bar \partial_{ij}}{ Z_{ij}}\:]\:\psi_2'
 = [\:E-N+\frac {\alpha}{2}N(N-1)\:]\:\psi_2'
\end{equation}
Clearly in the second case $\psi_2'$ will have to vanish at least
as fast as $\mid X \mid^{2\alpha}$.

Also, if $J\psi_1 = j_1 \psi_1$ , then for general $\beta, \gamma$
we have
$$J\psi_2 \:=\: [j_1 - \frac {N(N-1)}{2}\frac{\beta-\gamma}{2}]\psi_2
\:=\:j_2 \psi_2.$$
For the two cases noted above, $\beta=\gamma=\pm\alpha,\: j_2 = j_1$.
Incidentally if $\beta=\alpha,\: \gamma = -\alpha $ choice is made,
then
\begin{equation}
\psi_1 = (\frac{X}{\bar X})^{\alpha /2}\psi_2''
\end{equation}
and $\psi_2''$ satisfies the simple equation,
\begin{equation}
 [ -2 \partial_i \bar \partial_i +  Z_i \partial_i +
\bar Z_i \bar \partial_i ]\psi_2''
 = [E-N]\psi_2''.
\end{equation}
That is we recover the so called ``anyon gauge'' where the wave functions
are multivalued.  If we want to retain the singlevaluedness of the
wave function then our choices are limited to $\gamma=\beta + integer$.
The three-body interaction term vanishes if $\gamma = -\alpha or
\beta = \alpha $.  Allowing a non-zero integer only changes the angular
momentum of the wave function with a proportionate change in the
energy.  We will, to begin with retain the angular momentum also
and thus have $\beta =\gamma =\pm \alpha$ as the two choices.

At this stage we can summarise the known exact solution in a
convenient way \cite{exact}.  With the choice $\beta = \gamma =\alpha,
\psi_1=\mid X \mid^\alpha \psi_2$, we have two
basics types of exact solutions:
\begin{itemize}
\item
\begin{equation}
\psi_2 = \psi_2(\bar Z_i) ;\hspace{0.5cm} j < 0
\end{equation}
with the energy eigenvalues given by,
\begin{equation}
E = N - j + \alpha \frac {N(N-1)}{2}
\end{equation}
and
\item
\begin{equation}
\psi_2 = \psi_2(t=\sum_i\mid Z_i\mid^2)\; ;\hspace{0.5cm} j = 0.
\end{equation}
with $\psi_2(t)$ a polynomial of degree m in t.
The corresponding energy eigenvalues are given by,
\begin{equation}
E = N + 2m + \alpha \frac {N(N-1)}{2},
\end{equation}
The second solution is necessarily bosonic since $t$ is symmetric,
where as the first solution
needs explicit symmetrization and antisymmetrization of the wavefunction
in terms of $\bar Z_i$ to obtain the bosonic and fermionic wavefunctions.
Since this is always possible, the degeneracy of the first type solution
is exactly the same for both bosonic and fermionic type solutions for any
given angular momentum j ($ < 0 $);

Obtaining the first solution is trivial since only the ``scaling
operator'' contributes and $-j$ in the energy eigenvalue is nothing
but the degree of scaling.  The second solution can be obtained
most easily by transforming the equation in the form of a confluent
hypergeometric equation:
\begin{equation}
[t\frac{d^2}{dt^2}+(b-t)\frac{d }{dt}-a]\psi_2(t)=0,
\end{equation}
where
$$a=-\frac{1}{2}[E-N-\alpha \frac{N(N-1)}{2}] $$
$$b=N+\alpha\frac{N(N-1)}{2} $$
The admissible solutions are the regular Confluent Hypergeometric
Functions(CHF), $M(a,b,t)$ .  That $a = -m$ follows from the demand
of normalizability of the wave function. The corresponding eigenvalues follow
immediately.

We may also take a combination (product) of the solutions the types
discussed above, Eqns(21) and (23), to get further $j<0$ solutions. These
are :
\begin{equation}
\psi_2 = f(\bar Z_i)g(t); \hspace{1cm} E_{m,j}=2m - j +N +\alpha
\frac{N(N-1)}{2}.
\end{equation}
\item
With the choice $ \beta=\gamma=-\alpha$ and $ \psi_1 =\mid X \mid^
{-\alpha}\:\psi_2$ we have only the analouge of the first type of solutions
i.e. $\psi_2 = \psi_2(Z_i) $. $\psi_2$ has to be of the form $X^{d}
f(Z_i)$ with $d \ge 2 \alpha $
so that $\psi_1$ vanishes if
the coordinates of any two particles coincide. This means that j has to
be large enough($j > (N-1)(N-2)/2$).
The corresponding eigenvalues are given by
\begin{equation}
E_{j}= j +N -\alpha \frac{N(N-1)}{2}.
\end{equation}
If however the condition, $j > (N-1)(N-2)/2$, is not
satisfied then the wave function
$\psi_1$ remains regular only for some values of $\alpha$
($0\le \alpha \le 2j/N(N-1)$)
but not for all $0 \le \alpha \le 1$ which gives rise to the so called
noninterpolating solutions which have also been discussed in the
literature\cite {mlbd}.
To the best of our knowledge these are all the exactly known solutions
for the N anyon problem.
\end{itemize}
All these solutions for the energy eigenvalues have a linear
dependence on $\alpha$ while the corresponding eigenfunctions
are finite order polynomials apart from the $\mid X \mid^\alpha$
and the Gaussian factor.  In fact a simple scaling arguement shows
that if an exact eigenfunction is a polynomial (i.e. has a finite
degree), then
the corresponding eigenvalue must be linear in $\alpha$.

For, if $\psi_2$ is a polynomial in $Z_i,\bar Z_i$, with highest
total degree d then for $Z_i,\bar Z_i \rightarrow \infty$ the
polynomial becomes a monomial and only the scaling operator term
dominates giving,
\begin{equation}
E=N+d+\alpha \frac {N(N-1)}{2}.
\end{equation}
Now if $\mid Z_{ij}\mid \rightarrow 0$ every eigenfunction vanishes
as $\mid Z_{ij} \mid^\alpha$ or faster, i.e., after $\mid X \mid ^\alpha$
is taken out the remaining function has only integer powers of
$Z_i, \bar Z_i$.  Thus d has to be an integer and hence E has a
specific linear dependence on $\alpha$.

However it is well known that solutions which have nonlinear dependence
exist. This has been shown numerically \cite{numeral}, through perturbative
analysis \cite{perturb} for three anyon problem and using meanfield methods for
large number anyons \cite{tf}. This point is also our focus from now
on. The scaling arguement will then imply that d has to be
a nonlinear function of $\alpha$. How is this achieved?

Suppose $\psi_2$ admits a power series representation with
infinite radius of convergence but the series does not truncate
then the scaling arguement fails.  However the analysis of such a
series solution always seems to lead to exponentially divergent
behaviour making the solution non-normalizable, ie., the power
series has to truncate.  But then only linear soutions are possible.
The other possibility then is that the power series has
a finite radius of convergence, i.e., a ``scale'' R has to enter
if nonlinearity is to be possible.  But this means that
$\psi_2$ has two different representations as the scale parameter
$\lambda \rightarrow 0 $ and $\lambda \rightarrow \infty$, each
being valid for $\lambda < \lambda_{max}$ and $\lambda > \lambda_{min}$
respectively.  For $\lambda \ge \lambda_{min}$ the scaling
arguement can still work but now the series need not have only
integer powers, i.e., d can be a nonlinear function of $\alpha$.  We
conclude then that if a nonlinearly interpolating eigenvalue is
possible at all, the corresponding eigenfunction must have
two different series representations for $\psi_2(\lambda \hat Z_i,
\lambda \hat {\bar Z_i})$ as $\lambda \rightarrow 0$ and $\lambda
\rightarrow \infty$.  One then has to try to match the two series
suitably.

Exploring this possibility looks extremely complicated
for general N.  So we now specialise to N=3 and go over to relative
coordinate dynamics to reduce the number of independent variables
in the differential equation. These are given interms of the
Jacobi coordinates,
\begin{eqnarray}
\rho_1 = \frac {Z_1 -Z_2}{\sqrt 2} \\
\rho_2 = \frac {Z_1 +Z_2 -2 Z_3}{\sqrt 6}
\end{eqnarray}
and similarly their complex conjugates.

After separating out the CM coordinate and using the notation
$\rho_i = r_i exp(i\theta_i)$ the relevant hamiltonion
becomes,
\begin{eqnarray*}
H_{rel} & = &-\frac{1}{2}\sum_{i=1,2}\: [ \: \frac {\partial^2}{\partial r_i^2}
+
\frac {1}{r_i} \frac{\partial}{\partial r_i} +
\frac {1}{r_i^2} \frac{\partial^2}{\partial \theta_i^2}\: ] +
r_i \frac{\partial}{\partial r_i} \\
 & & -\frac{3\alpha}{3r_2^2-r_1^2 e^{2i(
\theta_2-\theta_1)}}\; [\frac{r_2^2-r_1^2 e^{2i(\theta_2-\theta_1)}}{r_1}
( \frac{\partial}{\partial r_1} -
\frac {i}{r_1} \frac{\partial}{\partial \theta_1}) \\
 & & +2r_2
( \frac{\partial}{\partial r_2} -
\frac {i}{r_2} \frac{\partial}{\partial \theta_2})\; ]
\end{eqnarray*}
The full wave function is ofcourse,
\begin{equation}
\psi = \mid X \mid^{\alpha} exp( -(1/2)\sum_i r_i^2)\phi (r_1,r_2,
\theta_1, \theta_2) \Phi_{CM} (R,\Theta).
\end{equation}

So far two asymptotic regions have been used to simplify the original
eigenvalue equation:
\begin{flushleft}
(i)$\hspace{0.3cm}
r_1,r_2 \rightarrow \infty$ such that $r_1/r_2$ is fixed and finite.\\
(ii)$\hspace{0.3cm}
 r_1,r_2 \rightarrow 0 $ such that $r_1/r_2$ is fixed and finite.
\end{flushleft}
and we arrived at the possibility of two different series for
some wave functions.  Clearly there are two more regions
possible, namely,
\begin{flushleft}
(iii)$\hspace{0.3cm}
r_1 \rightarrow 0$ such that $r_2$ is fixed or
$r_2 \rightarrow \infty $ such that $r_1$ is fixed,i.e.,$r_1/r_2
\rightarrow 0$.\\
(iv)$\hspace{0.3cm}
r_2 \rightarrow 0$ such that $r_1$ is fixed or
$r_1 \rightarrow \infty $ such that $r_2$ is fixed,i.e.,$r_2/r_1
\rightarrow 0$.
\end{flushleft}
Notice that these types of regions become possible only for $N \ge 3$,
when there are at least two relative coordinates.  For two anyon
problem there is a single relative coordinate and asymptotics for $\phi$
can be considered only as $r \rightarrow \infty$ or $r\rightarrow
0$ (regions (i) and (ii)).  Since two anyons have no possibility of any
finite R or scale parameter entering one has to use Taylor series,
and hence the wave functions are necessarily polynomials. This leads to
 a linearly
interpolating spectrum.  This is true provided either that $\alpha$
dependence is assumed to be smooth or $\phi$ is $C^{\infty}$ function.

To focus, let us consider (iii).  As $r_1/r_2 \rightarrow 0$ the
Hamiltonian becomes separable and $J_i = -i \frac {\partial}{\partial
 \theta_i}$ commute with the corresponding separated Hamiltonian.
This is true under the assumption that the derivatives in the $\alpha$
dependent terms in the $H_{rel}$ remain finite as $r_1/r_2 \rightarrow 0
$.
\begin{equation}
H_i = -\frac{1}{2}\left [\frac {\partial^2}{\partial r_i^2} +
\frac {1}{r_i} \frac{\partial}{\partial r_i} +
\frac {1}{r_i^2} \frac{\partial^2}{\partial \theta_i^2}\right ] +
r_i \frac{\partial}{\partial r_i} - \frac {\beta_i}{r_i}
\left ( \frac{\partial}{\partial r_i} -
\frac {i}{r_i} \frac{\partial}{\partial \theta_i}\right );\: i=1,2,
\end{equation}
where
$$\beta_1 = \alpha ;\hspace{0.3cm}  \beta_2 = 2\alpha$$
and
$$H = H_1 +H_2.$$
That $\beta_1=\alpha$ is obvious because by definition $r_1$ involves
particles 1 and 2.  $\beta_2=2\alpha$ follows from the fact that
$r_2$ involves all the three particles, and in the asymptotic region
particle 3 can only wind around particles 1 and 2 together, but not
around any one of them seperately.

Because of this as $r_1/r_2 \rightarrow 0$,
\begin{equation}
\phi(r_1,r_2,\theta_1,\theta_2) \;
\longrightarrow \;
 \phi_1(r_1,\theta_1) \phi_2(r_2,\theta_2),
\end{equation}
where,
$$\phi_i(r_i,\theta_i) =\eta_i(r_i) exp(in_i\theta_i); \; n_1+n_2 =j. $$
Now consider $r_1/r_2 \rightarrow 0$ as $r_1\rightarrow 0$ and
$r_2$ finite and fixed.  Let,
\begin{equation}
\phi(r_1,r_2,\theta_1,\theta_2) \;
\longrightarrow \;
r_1^\lambda exp(in_1\theta_1) \phi_2(r_2,\theta_2).
\end{equation}
Since H is seperable, we may first compute $H_1$ acting on $\phi$.
We have,
\begin{equation}
H_1\phi = r_1^\lambda exp(in_1\theta_1)\phi_2(r_2,\theta_2)
[\lambda - \frac {1}{2} \frac{(\lambda^2 - n_1^2 + 2\alpha (n_1
+\lambda))}{r_1^2}].
\end{equation}
Because $r_1 \rightarrow 0$ limit has to be finite (leading power
is already taken out) we get,
$$\lambda^2 - n_1^2 + 2\alpha (n_1+\lambda)=0 \Longrightarrow
\lambda = -\alpha \pm \mid n_1 -\alpha \mid $$
Now in the same limit , we have $\mid X\mid ^\alpha \rightarrow
r_1^\alpha r_2^{2\alpha}$ and therefore the total power
of $r_1$ becomes $\pm \mid n_1 -\alpha \mid$.  Since $\phi$ has to
vanish as $r_1 \rightarrow 0$ ($r_2$ fixed) we have to choose the
positive sign.

This also matches with the two anyon wave function in relative
coordinates which vanishes as $r^{\mid n - \alpha \mid }$.  Thus
choosing the positive sign,
$$\lambda =-\alpha +\mid n_1 -\alpha \mid $$
we have
\begin{eqnarray}
(H_1+H_2)\phi& =& [(H_2+\lambda)\phi_2] r_1^\lambda exp(in_1\theta_1)\\
 & = & (E -2 -3\alpha)\phi_2 r_1^\lambda exp(in_1\theta_1)
\end{eqnarray}
Thus $\phi_2$ satisfies the eigenvalue equation
\begin{equation}
H_2 \phi_2 = (E-2-3\alpha -\lambda)\phi_2 =(E-2-2\alpha -\mid n_1
-\alpha \mid)\phi_2 ={\cal E} \phi_2.
\end{equation}
$H_2$ is once again the ``two anyon Hamiltonian''.  There is one crucial
difference; since our equations are valid only asymptotically $r_1/r_2
\rightarrow 0$ we have to maintain $r_2$ nonzero. Let us say that
$r_2 \ge R_0$, where $R_0$ is some new parameter.

With the definitions
\begin{eqnarray}
\mu & = & \frac {1}{2} [\:-2\alpha + \mid n_2 -2\alpha\mid \:]\\
 b  & = & 1+\mid n_2 -2\alpha\mid \\
 a  & = & -\frac {1}{2}[\;{\cal E} +2\alpha -\mid n_2 -2\alpha \mid \;] \\
    & = & -\frac{1}{2}[\;E - 2 -\mid n_1 -\alpha \mid \;-\mid n_2 -
2\alpha \mid \;] \\
 x  & = & r_2^2 \\
 R  & = & R_0^2 \\
 \phi_2(r_2,\theta_2)& = &x^\mu exp(in_2\theta_2) u(x)
\end{eqnarray}
the eigenvalue equation for $\phi_2$ becomes the CFE for u(x):
\begin{equation}
[x\frac{d^2}{dx^2}+(b-x)\frac{d }{dx}-a]u(x)=0,
\end{equation}
$$R \le x < \infty.$$
The general solution is given by,
\begin{equation}
u(x) = C_1 M(a,b,x) +C_2 U(a,b,x),
\end{equation}
where M and U are in standard notation \cite{cfe} and  $C_1$ and $C_2 $
are constants.

For $C_1$ we have two possibilities, namely it is
zero or non zero. If it is non zero then normalizability as $x \rightarrow
 \infty$ requires a to be minus a non negative integer. This is
precisely the case when the U solution ceases to be an independent
solution. In other words, if $C_1$ is non zero then we really have only
one solution which is a finite order polynomial. This solution is also
well behaved as x goes to zero and therefore we may take R to be zero.
The energy is read off
by putting the value of a in the definitions given above. We see that
the energy eigenvalue in this case is linear in $\alpha$. The known
exact solutions are covered by this case.

If $C_1$ is zero then $C_2$ is non zero and we have only the U solution.
U is well behaved for large x and thus we do not get any condition on a
from the demand of normalizability. However U is divergent as $x
\rightarrow 0$ and we have to keep R strictly positive.

The only way E can get quantised now is by putting a boundary condition
at $x =R$.  Hermiticity analysis implies $ U\mid_R =0 $ or $dU/dx
\mid_R = 0$. But
$$ \frac {dU}{dx} = -a U(a+1,b+1,x) $$
Because a is not a non positive integer (else we are back to the
previous case ) essentially we get the conditions
$$ U(a,b,x) =0 $$ or $$ U(a+1,b+1,x)=0.$$ Now U(a,b,x) has real
positive zeros only if \cite{cfe}
$$ a<0 \hspace{0.5cm} and \hspace {0.5cm} 1+a-b <0.$$
In the present case $b=1+\mid n_2 -2\alpha\mid $ and
$$1+a -b =a\: -\mid n_2 -2\alpha \mid$$
Zeros of U exist only if,
\begin{eqnarray*}
a < 0 \hspace{0.5cm} & and & \hspace{0.5cm} a < \; \mid n_2 -2\alpha \mid
\hspace{0.5cm} (if \; U = 0)\\
& or & \\
a+1 < 0 \hspace{0.5cm} & and & \hspace{0.5cm} a < \; \mid n_2 -2\alpha \mid
\hspace{0.5cm} (if \; U' = 0),
\end{eqnarray*}
that is for $a<0 $ or $a<-1$ respectively.  In the limit $\alpha
\rightarrow 0 $ we know that all solutions are finite polynomials.
In effect then $R\rightarrow 0$ must hold as $\alpha \rightarrow
0$, i.e., $a$ must be a negative integer.  The Bosonic ground state
has energy 2$\hbar\omega$.  To admit this possibility we must
allow for $a$ to be zero.  We will therefore choose $U(a,b,R)=0$
as our boundary condition.  Since for $a<0$ (a not an integer)
U has in general more than one zero we will stipulate $R$
to be the smallest zero of U (say).  This then determines uniquely $a$
and hence E as a function of R which is still an unknown parameter.

It may seem that the above boundary condition is very restrictive.
It is not.  For suppose we knew $E=E(\alpha)$, then U has zeroes and
hence the smallest zero.  Therefore for all $\alpha$, we can always find
an R such that,
$U(a(\alpha),b,R)=0$.  But now clearly R depends on $\alpha$ and the
asymptotic quantum numbers $n_1,n_2$.  Thus $a(\alpha,n_1,n_2)$
determines $R=R(\alpha,n_1,n_2)$. Conversely, if we could
determine $R =R(\alpha,n_1,n_2)$ by some arguement then U = 0
will determine $E(\alpha,n_1,n_2)$.

The boundary  condition can thus be understood as stipulating that
there exists functions $a(\alpha), R(\alpha)$ and $b(\alpha)$ which is
explicitly known,
such that $U(a,b,R)=0$, i.e., in the space parametrised by variables E,
b and R with $2 \le E < \infty$,
$\: b=1\:+\:\mid n_2-2\alpha\mid,\: 0<R<\infty$ there exist curves along
which U=0.  Determining these curves is equivalent to getting the
spectrum. The differential statement of U = 0 along a curve is of
course ${dU\over {d\alpha}} = 0$ which can be written as:

\begin{equation}
{\partial U \over \partial a} {da \over d\alpha} +
{\partial U \over \partial b} {db \over d\alpha} +
{\partial U \over \partial R} {dR \over d\alpha}  = 0
\end{equation}

Meanwhile we have tested the U = 0 criterian in the following way.
We choose
a particular state at the fermionic or the bosonic end. Use the
numerically determined eigenvalue to get $a(\alpha$) and determine the
zero of the U function numerically. Thus we get a R($\alpha$). We fit a
form to thus determined R and the compute E at arbitrary values of
$\alpha$. This still has predictive power in that once we postulate a
form for R($\alpha$) we can determine its parameters for small
values of $\alpha$ where perturbation theory can be used. We then have a
prediction for all values of $\alpha$ in the range [0,1]. In particular
we have carried out this exercise for the state which interpolates to
the fermionic ground state and the corresponding bosonic state with
j=2 which
is supersymmetric \cite{numeral} to the fermionic ground state.
The results are given in the figure 1 for the energy and figure 2
for the corresponding R .

In the end we briefly mention another method of analysing solutions
in the three anyon problem using the Fourier expansion of the wavefunction
$\phi$ in eq.(31).  Note that $\phi$ may be written as,
\begin{equation}
\phi_j(r_1,r_2,\theta_1,\theta_2)=
\sum_{n_1,n_2} e^{i n_1\theta_1}e^{i n_2\theta_2}\chi_{n_1,n_2}(r_1,r_2)
=\sum_{n} e^{i j \frac{(\theta_1+\theta_2)}{2}}
e^{i n \frac{(\theta1-\theta_2)}{2}}\chi_{j,n}(r_1,r_2)
\end{equation}
for a given $j$, where $j=n_1+n_2$, $n=n_1-n_2$.  We can now use this
representation to look at the solutions for a given $j$. The eigenvalue
equation now becomes $H_{rel}\phi_j = \cal E \phi_j $, where $H_{rel}$
is the three-anyon Hamiltonian in relative coordinates and ${\cal E} =
E-2-3\alpha$ as before.  Substituting for $\phi_j$ the eigenvalue equation
takes the form,
\begin{eqnarray*}
& &3x_2[4x_1\frac{\partial^2}{\partial x_1^2}+4(1+\alpha-x_1)\frac{\partial}
{\partial x_1} -\frac{(j+n)(j+n-4\alpha)}{4x_1}\\
&+&4x_2\frac{\partial^2}{\partial x_2^2}+4(1+2\alpha-x_2)\frac{\partial}
{\partial x_2} -\frac{(j-n)(j-n-8\alpha)}{4x_2}+2{\cal E}]\chi_{j,n}\\
&=&x_1[4x_1\frac{\partial^2}{\partial x_1^2}+4(1+3\alpha-x_1)\frac{\partial}
{\partial x_1} -\frac{(j+n+4)(j+n+4-12\alpha)}{4x_1}\\
&+&4x_2\frac{\partial^2}{\partial x_2^2}+4(1-x_2)\frac{\partial}
{\partial x_2} -\frac{(j-n-4)^2}{4x_2}+2{\cal E}]\chi_{j,n+4},
\end{eqnarray*}
where $x_i=r_i^2$.  This equation is exact.  These infinite set of
coupled equations relate Fourier modes differing by 4, i.e., if
$\chi_0,\chi_1,\chi_2,\chi_3$ are know then $\chi_{4k},\chi_{4k+1},
\chi_{4k+2}, \chi_{4k+3}$ get determined in terms of $ \chi_0,...,\chi_3$
etc.  However there is no relation among $\chi_0,\chi_1,\chi_2,\chi_3$
themselves.  In a sense these four functions will give four independent
solutions of the eigenvalue equations.  We can then deal with a given
``tower'' separately and independently and this is true
for every given j.  Let us concentrate on one tower. Now three distinct
cases arise naturally:
\begin{flushleft}
(a) Only one member of the tower is nonzero, ie., $\chi_{j,n} = \chi_{j,m}
\delta_{n,m}$.

(b) Only a finite number of $\chi_{j,n}$'s are nonzero, ie., $\chi_{j,n}
=0~~ \forall ~~n \ge n_1~~~and~~~\forall ~~  n \le n_2$ with $ n_2 < n_1$.

(c) Infinitely many $\chi_{j,n}$'s are nonzero.  This gives raise to the
following cases:
\begin{eqnarray*}
\chi_{j,n}&=&0 ~~\forall~~ n\ge n_1 \\
\chi_{j,n}&=&0 ~~\forall~~ n\le n_2 \\
\chi_{j,n}&\ne& 0 ~~\forall~~ n.
\end{eqnarray*}
\end{flushleft}

Case (a) is simple to analyse and it reproduces the known exact
solutions.  In the  exact eigenvalue equation relating $\chi_{j,n}$
and $\chi_{j,n+4}$ given above we set lhs to zero which gives an
equation for the $\chi_{j,n}$.  But the rhs also gives rise to another
equation for the same $\chi_{j,n}$ when the coupling to $\chi_{j,n-4}$
is taken into account.  However the form of equations suggests that
the two equations are seperable but must be consistently solved.
The consistency conditions immediately yield the linear exact solutions
which are already outlined previously.  Cases (b) and (c) seem quite
complicated and nonlinearly interpolating states must be in one of these
cases. Consider the normalisation condition on the full wavefunction,
\begin{equation}
\parallel \psi_j \parallel^2 =
\sum_n \int_0^{\infty} dx_1 dx_2 |X|^{2\alpha} |\chi_{j,n}|^2
e^{-{x_1+x_2}} \equiv \sum_n C_n
\end{equation}
If $\psi_j$ is given by case (b) then there are only a finite number
of terms in the norm and if $C_n$'s are finite then $\psi_j$ is
normalisable.  Thus the quantisation condition for the nonlinear
states must arise by the demand that $C_n$ must be finite.  For the
case (c) even if all $C_n$'s are finite we may still get the wavefunction
to be infinite because the sum above may not converge and then the
quantisation condition would be the convergence of $\sum_n C_n$.

We have not at present analysed this scenario in detail.  But it
does appear that the case (b) is ruled out in which case the exact
solutions come from case (a) and the nonlinear solutions arise
from case (c).

\section{Summary and Discussion}

We have in this paper, analysed the problem of three anyons confined
in an oscillator potential.  The energy spectrum of three anyons
generically contains two types of interpolations as a function of the
statistical parameter $\alpha$.  The first type are the linearly
interpoating eigenvalues with $\Delta E = E(\alpha=1)-E(\alpha=0)=3$.
These solutions are easily generalised to N-anyon case where
$\Delta E = N(N-1)/2, N\ge 2$.  The second set consists of solutions
for eigenvalues which are in general non-linear functions of $\alpha$.
In this paper we have outlined an analytical method of obtaining
such solutions.  We exploit the known asymptotics to arrive at these
solutions.
Since the differential equation is valid in all regions
it is sufficient to specify the wavefunction asymptotically to obtain
the exact eigenvalues.

For N=3, there are two relative coordinates whose magnitudes
are given by $r_1, r_2$.  The asymptotic regions are then defined by,
\begin {itemize}
\item (a) $r_1,r_2 \rightarrow \infty$.
\item (b) $r_1,r_2 \rightarrow 0 $.
\item (c) $r_1 \rightarrow 0,\hspace {0.5cm} r_2 \; fixed;\hspace{0.5cm} or
\hspace{0.5cm} r_2 \rightarrow \infty, r_1 \; fixed$.
\item (d) $r_2\rightarrow 0,\hspace {0.5cm} r_1 \; fixed;\hspace{0.5cm} or
\hspace{0.5cm} r_1 \rightarrow \infty, r_2 \; fixed$.
\end{itemize}
Asymptotic region (a) immediately yields the Gaussian which
controls the behaviour of the wavefunction at infinity  and region
(b) yields the Laughlin-Wu factor which regulates the wave function
as two anyons approach each other. The asymptotic
region (c) (or (d)) is used to obtain the nonlinearly interpolating solutions.

There is a caveat to note: There may be exact eigenfunctions
which do not satisfy the
asymptotic equation (33). In this case our analysis can at best be
expected to give an approximation to the exact eigenvalues. In
the absense of a complete rigorous proof precluding this possiblity we
have to note it as an open one.

A few additional points are worth noting.

i) We have chosen R to be the smallest zero of U without any
justification. For low lying eigenvalues the U function has only one
zero so there is no ambiguity. For higher states though more zeros are
possible and their implications are to be explored and elaborated.

ii) Some qualitative properties such as a tower of
evenly spaced eigenvalues
for every non linearly interpolating eigenvalue \cite{tf} and
``supersymmetry" for a class of eigenstates \cite{numeral} have been
noted in the literature cited. Whether and how these fit into our
appraach is to be seen.

\begin{flushleft}
{\bf Acknowledgements}\\
We thank G.Baskaran, Diptiman Sen and P.N.Srikanth for
valuable discussions.
We also thank Patricia Monger for the Super Mongo software, which
helped us in figuring out the zeros of Confluent
Hypergeometric Functions.
\end{flushleft}

\newpage
\begin{flushleft}
{\bf Figure Captions}\\
\vspace{1cm}
Figure 1: Plot of the Fermionic ground state energy and the corresponding
Bosonic j = 2 state as a function of $\alpha$ obtained by using the
boundary condition at R.
The non-interacting Bosonic
state is at $\alpha=0$
and the non-interacting Fermionic state is at $\alpha=1$ and energy $E=4$
in dimensionless units.  The points shown along the Fermionic curve
are taken from the numerical calculations \cite{numeral}.\\
\vspace{1cm}
Figure 2: Shows the values of R as a function of $\alpha$ used in
computing the energies of the corresponding states. For the Fermionic
ground state $R \approx 0.405 \alpha (1-\alpha)$ obtaied from numerical
computation close to $\alpha=0 $.  The corresponding form for the
Bosonic state is $R\approx \alpha^{0.6}$.
\end{flushleft}

\end{document}